\documentclass[conference]{IEEEtran}
\IEEEoverridecommandlockouts
\usepackage{cite}
\usepackage{amsmath,amssymb,amsfonts}
\usepackage{algorithmic}
\usepackage{graphicx}
\usepackage{textcomp}
\usepackage{comment}
\usepackage{xcolor,balance}
\def\BibTeX{{\rm B\kern-.05em{\sc i\kern-.025em b}\kern-.08em
    T\kern-.1667em\lower.7ex\hbox{E}\kern-.125emX}}
\usepackage[nolist]{acronym}
\begin{acronym}[ACRONYM]
\acro{3GPP}{The 3rd Generation Partnership Project}
\acro{3D}{three dimensional}
\acro{5G}{fifth generation}
\acro{6G}{sixth generation}
\acro{AI}{artificial intelligence}
\acro{AR}{augmented reality}
\acro{AoA}{angle-of-arrival}
\acro{AoD}{angle-of-departure}
\acro{BS}{base station}
\acro{CF}{consumption factor}
\acro{CSI}{channel state information}
\acro{DL}{downlink}
\acro{D-MIMO}{distributed multiple-input multiple-output}
\acro{EER}{energy efficiency ratio}
\acro{EM}{electromagnetic}
\acro{GDOP}{geometric dilution of precision}
\acro{GNSS}{global navigation satellite system}
\acro{HMI}{human-machine interaction}
\acro{IQ}{in-phase and quadrature}
\acro{JCS}{Joint Communication and Sensing}
\acro{ISAC}{integrated sensing and communication}
\acro{JRC}{joint radar and communication}
\acro{JRC2LS}{joint radar communication, computation, localization, and sensing}
\acro{ICI}{inter-carrier interference}
\acro{IOO}{indoor open office}
\acro{IoT}{Internet of Things}
\acro{IRN}{infrastructure reference node}
\acro{ISG}{industry specification group}
\acro{JCAS}{joint communications and sensing}
\acro{KPI}{key performance indicator}
\acro{KVI}{key value indicator}
\acro{LCA}{life cycle analysis}
\acrodefplural{LCA}{life cycle analyses}
\acro{LoS}{line-of-sight}
\acro{MAC}{medium access control}
\acro{MIMO}{multiple-input multiple-output}
\acro{ML}{machine learning}
\acro{mmWave}{millimeter-wave}
\acro{NF}{network function}
\acro{NLoS}{non-line-of-sight}
\acro{NR}{new radio}
\acro{NTN}{non-terrestrial network}
\acro{OFDM}{orthogonal frequency-division multiplexing}
\acro{OTFS}{orthogonal time-frequency-space}
\acro{PRS}{positioning reference signal}
\acro{PHY}{physical layer}
\acro{QoS}{Quality of Service}
\acro{RAN}{radio access network}
\acro{RAT}{radio access technology}
\acro{RedCap}{reduced capacity}
\acro{RF}{radio frequency}
\acro{RIS}{reconfigurable intelligent surface}
\acro{RTK}{real-time kinematic}
\acro{RTT}{round-trip-time}
\acro{SDG}{sustainable development goal} 
\acro{SLAM}{simultaneous localization and mapping}
\acro{SNR}{signal-to-noise ratio}
\acro{SIT}{sustainability, inclusiveness, and trustworthiness}
\acro{SOTA}{state of the art}
\acro{SL}{sidelink}
\acro{ToA}{time-of-arrival}
\acro{TDoA}{time-difference-of-arrival}
\acro{TR}{time-reversal}
\acro{TRP}{transmission and reception point}
\acro{TXRX}[TX/RX]{transmitter/receiver}
\acro{TX}{transmitter}
\acro{RX}{receiver}
\acro{UE}{user equipment}
\acro{UN}{United Nations}
\acro{multi-RTT}{multi-cell round-trip-time}
\acro{UL}{uplink}
\acro{UL-TDOA}{uplink time-difference-of-arrival}
\acro{DL-TDOA}{downlink time-difference-of-arrival}
\acro{UMi}{3D-urban micro}
\acro{UMa}{3D-urban macro}
\acro{UWB}{ultra-wide band}
\acro{FR1}{frequency range 1}
\acro{FR2}{frequency range 2}
\acro{WLAN}{wireless local-area network }

\end{acronym}

\IEEEoverridecommandlockouts \IEEEpubid{\makebox[\columnwidth]{978-8-3503-8544-1/24/\$31.00~\copyright{}2024 IEEE \hfill} \hspace{\columnsep}\makebox[\columnwidth]{ }}
    
\begin{document}
\bstctlcite{IEEEexample:BSTcontrol}
\title{Joint Communication and Sensing for 6G -- A Cross-Layer Perspective
\thanks{This work is partially supported by the European Commission through the Horizon Europe/JU SNS project Hexa-X-II (Grant Agreement no. 101095759).}
}
\author{%
Henk Wymeersch\IEEEauthorrefmark{1}, 
Sharief Saleh\IEEEauthorrefmark{1}, 
Ahmad Nimr\IEEEauthorrefmark{2}, 
Rreze Halili\IEEEauthorrefmark{3},
Rafael Berkvens\IEEEauthorrefmark{3},
Mohammad H. Moghaddam\IEEEauthorrefmark{4}, 
\\
José Miguel Mateos-Ramos\IEEEauthorrefmark{1}, Athanasios Stavridis\IEEEauthorrefmark{5},
Stefan Wänstedt\IEEEauthorrefmark{5},
Sokratis Barmpounakis\IEEEauthorrefmark{7},
Basuki Priyanto\IEEEauthorrefmark{6},
\\
Martin Beale\IEEEauthorrefmark{6}, Jaap van de Beek\IEEEauthorrefmark{8}, Zi Ye\IEEEauthorrefmark{8}, Marvin Manalastas\IEEEauthorrefmark{9}, Apostolos Kousaridas\IEEEauthorrefmark{10},
Gerhard P. Fettweis\IEEEauthorrefmark{2} 
\\
\footnotesize
\IEEEauthorrefmark{1}Chalmers University of Technology, Sweden, 
\IEEEauthorrefmark{2}TU Dresden, Germany,
\IEEEauthorrefmark{3}IDLab, imec-UAntwerp, Belgium,\\
\IEEEauthorrefmark{4}Qamcom Research and Technology, Sweden
\IEEEauthorrefmark{5}Ericsson Research, Sweden,\\
\IEEEauthorrefmark{6}Sony Europe, Sweden
\IEEEauthorrefmark{7}WINGS ICT Solutions, Athens, Greece
\IEEEauthorrefmark{8}Lule\aa\ University of Technology, Sweden, 
\IEEEauthorrefmark{9}Nokia, Finland
\IEEEauthorrefmark{10}Nokia, Germany
}

\maketitle

\begin{abstract}
As 6G emerges, cellular systems are envisioned to integrate sensing with communication capabilities, leading to multi-faceted communication and sensing (JCAS). 
This paper presents a comprehensive cross-layer overview of the Hexa-X-II project's endeavors in JCAS, aligning 6G use cases with service requirements and pinpointing distinct scenarios that bridge communication and sensing. 
This work relates to these scenarios through the lens of the cross-layer physical and networking domains, covering models, deployments, resource allocation, storage challenges, computational constraints, interfaces, and innovative functions.  
\end{abstract}


\vspace{-5mm}
\section{Introduction}

6G is expected to be the first generation of cellular systems with built-in sensing capabilities \cite{liu2022integrated}. The inclusion of sensing capabilities within cellular systems 
opens many opportunities, but also challenges. 
Challenges include not only the needed technical solutions but also identifying the best usage and providing privacy and security. 
Sensing is a broad concept in communication systems \cite{du2022overview}, including radar-like sensing (locating objects within the coverage of a radio network, that are not necessarily connected to the network), imaging and spectroscopy, as well as general feature extraction from received waveforms \cite{hexax_d33}.  Conventional positioning  (i.e., locating \acp{UE}) can thus be interpreted as a service relying on sensing. 

\begin{figure}
    \centering
    \includegraphics[width=1\columnwidth]{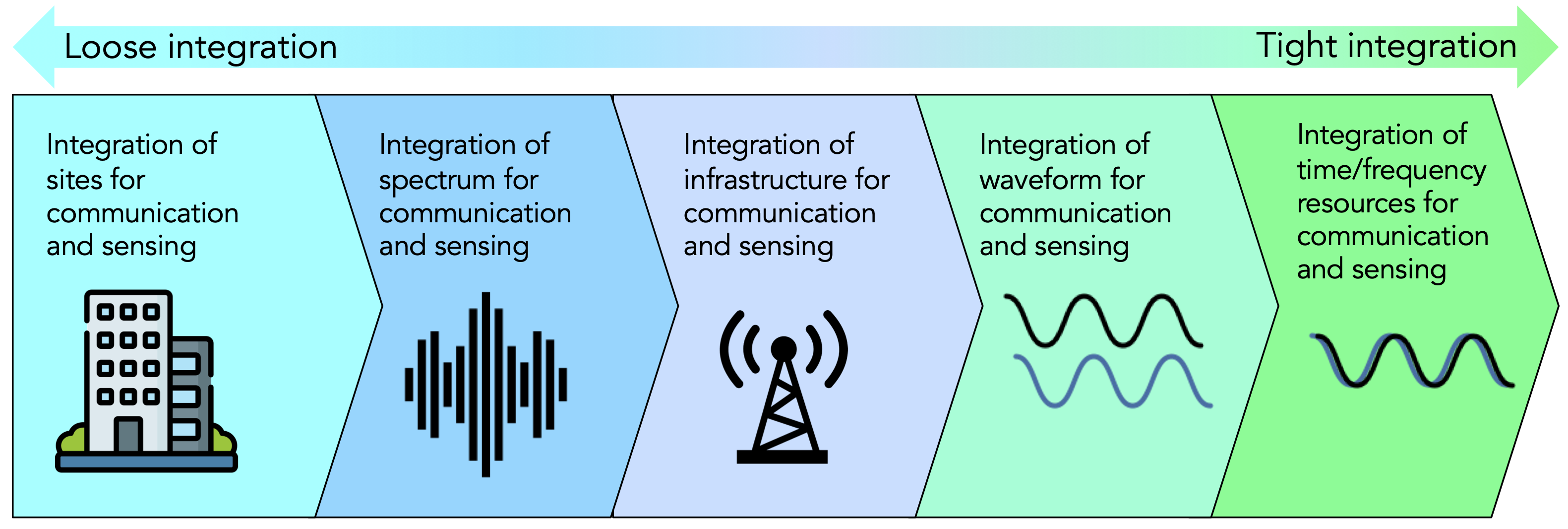}
    \caption{\ac{JCAS} levels of integration: from very loose to very tight integration, ranging from integration of sites, integration of spectrum, integration of infrastructure, and integration of waveforms, to integration of radio resources. } \vspace{-5mm}
    \label{fig:ISACIntegration}
\end{figure}

The use of communication signals for sensing and the use of sensing information for improving communication leads to the concepts of \ac{ISAC}
and \ac{JCAS}, which we treat as equivalent,\footnote{Strictly speaking, \emph{joint} means 'shared, held, or made by two or more parts together', while \emph{integrated} means 'with various parts or aspects linked or coordinated'. We will consider these concepts interchangeable and therefore use only \ac{JCAS}.} and
aim to endow the communication system with sensing capabilities. 
\ac{JCAS} can be broken down in different ways, based on the level of integration, as shown in Fig.~\ref{fig:ISACIntegration}. On the left side, very loose integration is shown, where sensing is only integrated in terms of the communication sites, but is based on an external sensing system (e.g., camera) \cite{alkhateeb2023deepsense} that provided information to improve communication, while on the right side, very tight integration is shown, where sensing is concurrent with data transmission, using the same radio resources \cite{keskin2021limited}. All levels of integration involve trade-offs and synergies between communication and sensing \cite{liu2022integrated}.  

There has been a large number of research papers on \ac{JCAS}, ranging from broad visions  \cite{demirhan2023integrated,wei2022toward,behravan2022positioning} to detailed technical contributions \cite{xiong2023fundamental,keskin2021limited} as well as extensive survey and overview papers \cite{liu2022integrated,wei2023integrated,zhang2021enabling,liu2022survey}. In parallel, demonstration of \ac{JCAS} has been covered in, e.g.,  \cite{ji2023networking,wang2023road,khatib2023designing,callebaut2022techtile}.  Nevertheless, few works have addressed  \ac{JCAS} from a broader perspective in the 6G context, considering the use cases and applications, the relation between the physical and networking layers, and the sustainability and trustworthiness aspects.  The EU Flagship Projects Hexa-X (https://hexa-x.eu) and Hexa-X-II (https://hexa-x-ii.eu) aim to address this broader perspective, considering 6G radio from its many facets and thereby supporting networks beyond communication in a way that meets the values of our future society \cite{hexaxii_d11}. 

In parallel, work is ongoing in standardization, including in  3GPP, {IEEE}, and ETSI. 
As a global cellular telecommunications standardization body, 3GPP foresees multiple market segments and verticals where 5G-based sensing services can be beneficial for intelligent transportation, aviation, enterprise, smart city, smart home, factories, consumer applications, telepresence, and the public sector \cite{sa1sid}. 3GPP has investigated the potential of integrated sensing and communication technology for enabling new services and use cases for various applications \cite{sa1tr}. 
The work on radio-related aspects is likely to start by extending the existing channel models in \cite{ra1tr} to include the characterization of a sensing-related channel model, incorporating aspects related to the reflection and scattering of signals. Subsequent work, in future 3GPP releases, would then consider technical solutions to support JCAS in terms of the system architecture and radio access network. 
In addition, in IEEE 802.11, the task group BF (TGBF) is responsible for the introduction of \ac{WLAN} sensing \cite{restuccia2021ieee}. 
More specifically, TGBF aims to define the use of the \ac{PHY} and \ac{MAC} features of the IEEE 802.11 standard for extracting measurements that can characterize features, such as the range and velocity, of objects in an area of interest.  
The work in these standardization bodies focuses on the near future, while Hexa-X and Hexa-X-II consider a longer time horizon.

In this paper, we aim to provide an overview of the work in Hexa-X-II as it pertains to \ac{JCAS}. Starting from 6G use cases, service requirements are extracted, leading to a small set of more precise scenarios, for both communication and \ac{JCAS}. Relevant \acp{KPI} are defined and the need for novel \acp{KPI} related to sustainability and trustworthiness is emphasized. Next, the scenarios are considered from the perspective of the physical and networking layers, considering models, deployments, resource allocation, and signal processing methods, and then storage and computational challenges, interfaces, exposure, and new functions.

\section{From \ac{JCAS} Use Cases and \acp{KPI} to a Cross-Layer Perspective}

In this section, the focus is on 6G use cases and their relation to \ac{JCAS}. These use cases are distilled into four scenarios, for which new \acp{KPI} are introduced. 
Finally, the need for a cross-layer perspective is articulated.

\subsection{Exemplifying  Use Case}

In traffic safety, \ac{JCAS} can enhance vehicle sensing range, especially in urban settings. For instance, at a blind corner, a vehicle could request network sensing for pedestrians or other vehicles nearby. Even when a base station with \ac{LoS} is absent, sensing-capable \acp{UE}  would cooperatively collect and send the necessary data to the vehicle, aiding the driver in making safer turning decisions.


\begin{figure}
    \centering
    \includegraphics[width=1\columnwidth]{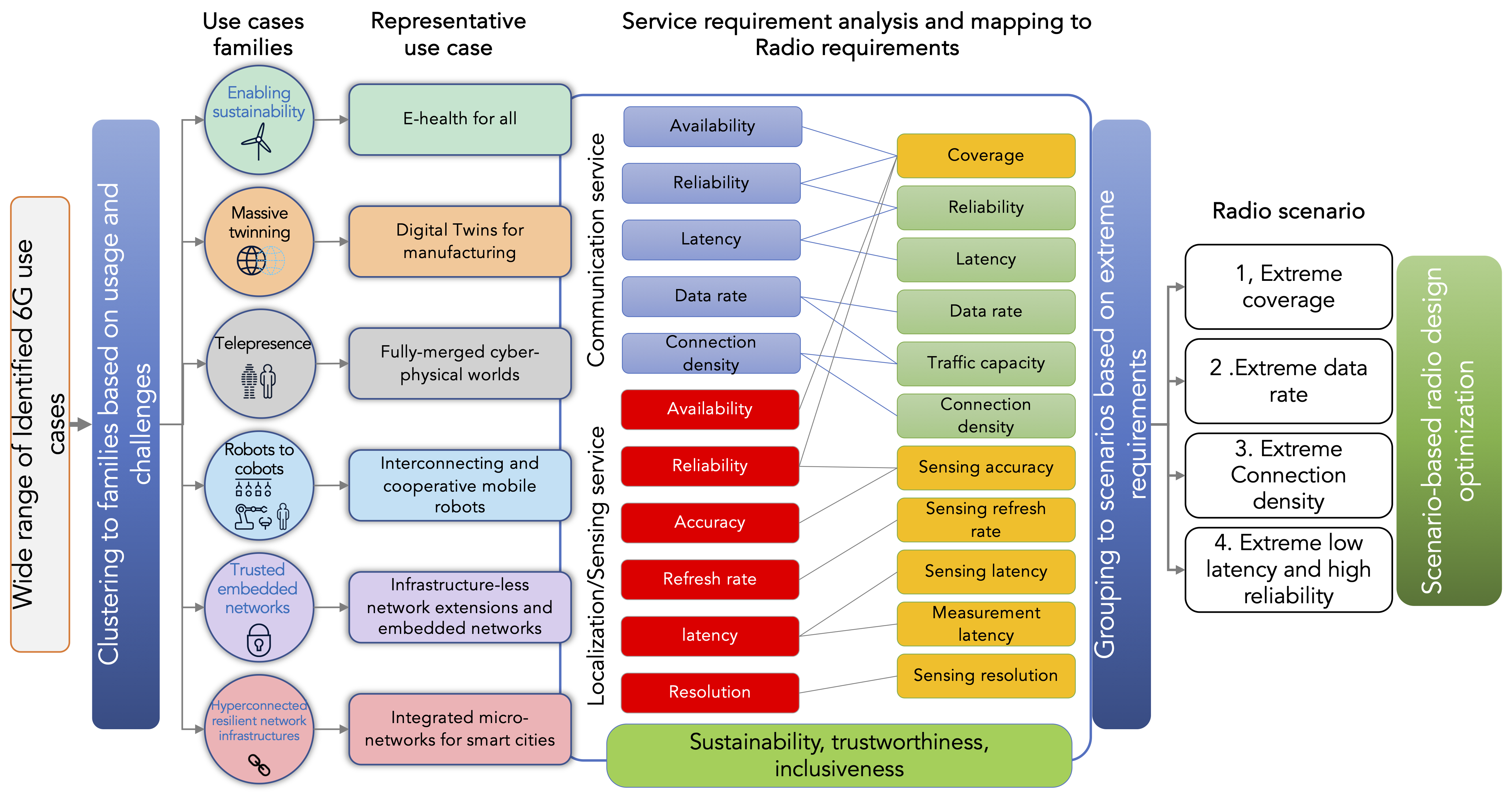}
    \caption{JCAS within the 6G use cases and radio scenarios, as well as \acp{KPI}. Zoom in for details. The figure illustrates a framework for analyzing the radio requirements of use cases. A use case is characterized by a set of service requirement attributes, which are associated with a set of radio requirement attributes.  Based on the analysis of six representative use cases, four radio scenarios emerged, each emphasizing extreme values in a particular radio requirement attribute.}\vspace{-5mm}
    \label{fig:scenarios}
\end{figure}
\subsection{From Use Cases to Radio Scenarios}\label{sec:radioscenarios}

Hexa-X has introduced a set of use case families (see Fig.~\ref{fig:scenarios}) via the combination of a top-down approach driven by vision, core values, and research challenges, and a bottom-up approach to illustrate the use of new technologies. 
It has been observed that the Hexa-X-II use cases as shown in Fig.~\ref{fig:scenarios} have also been part of the study in 3GPP \cite{sa1sid}:  smart cities, e-health, and mobile robots are among the use-cases. 

Based on analyses and requirements of these use case families \cite{hexax_d31}, Hexa-X-II has defined a small set of radio scenarios, with well-defined \ac{KPI} intervals.
\begin{enumerate}
    \item \emph{Extreme data rate:} with rates above 10 Gb/s, relying on wide bandwidths and localized coverage. Despite large bandwidths and the potential for accurate positioning and sensing, the requirements for positioning and sensing are expected to be very loose. 
    \item \emph{Extreme low latency and high reliability:} with lower rates, but stringent requirements on positioning and sensing. 
    \item \emph{Extreme connection density:} for medium data rates, and reduced requirements on positioning and sensing.
    \item  \emph{Extreme coverage:}  providing communication, positioning, and sensing services everywhere with variable requirements.
\end{enumerate}
\ac{JCAS} will play an important role in all the above scenarios, considering the expected developments on enabling technologies on 6G radio, networking, and intelligent computation and protocols. Further details are elaborated in Sections III and IV.

\subsection{New \acp{KPI} in Support of the Radio Scenarios}
The benefit of the four radio scenarios is that they allow the definition of \ac{KPI} ranges. This includes the conventional \acp{KPI} for positioning and sensing (such as accuracy, resolution, latency, availability), which are elaborated, e.g., in \cite{behravan2022positioning}. On the other hand, as one of the 6G cornerstones will be improving sustainability and trustworthiness, new measurable \acp{KPI} must be introduced \cite{wymeersch20236g}. It is noteworthy that trustworthiness and sustainability, as one of the key aspects in Hexa-X-II, have also been considered in \cite{sa1sid}.

\subsubsection{Sustainable \ac{JCAS}}
As an example, power consumption holds significant importance for designing and optimizing \ac{JCAS}, particularly in achieving a balance between sensing and communication performance due to shared resources \cite{10000730}.
Within Hexa-X-II, to analyze the expected shifts in power consumption, two metrics are considered. 
    First, \emph{the \ac{EER}} is based on \cite{murdock2013consumption} and captures the ratio between the output power (in Watts) and the total power consumed (also in Watts) by the sensing or communication system (or a component thereof) to achieve this power output. 
    Second, \emph{the \ac{CF}} extends the ideas from  \cite{murdock2013consumption}, where the \ac{CF} captures the ratio between the maximum rate (in bits/s) and the total power consumed (in Watts) for communication. 
    In positioning and sensing, in place of the maximum rate, we consider the detection probability, ranging accuracy, angle estimation accuracy, correct classification rate, or radar mutual information, to define the \ac{CF}.
\begin{figure}
    \centering
    \includegraphics[width=1\linewidth]{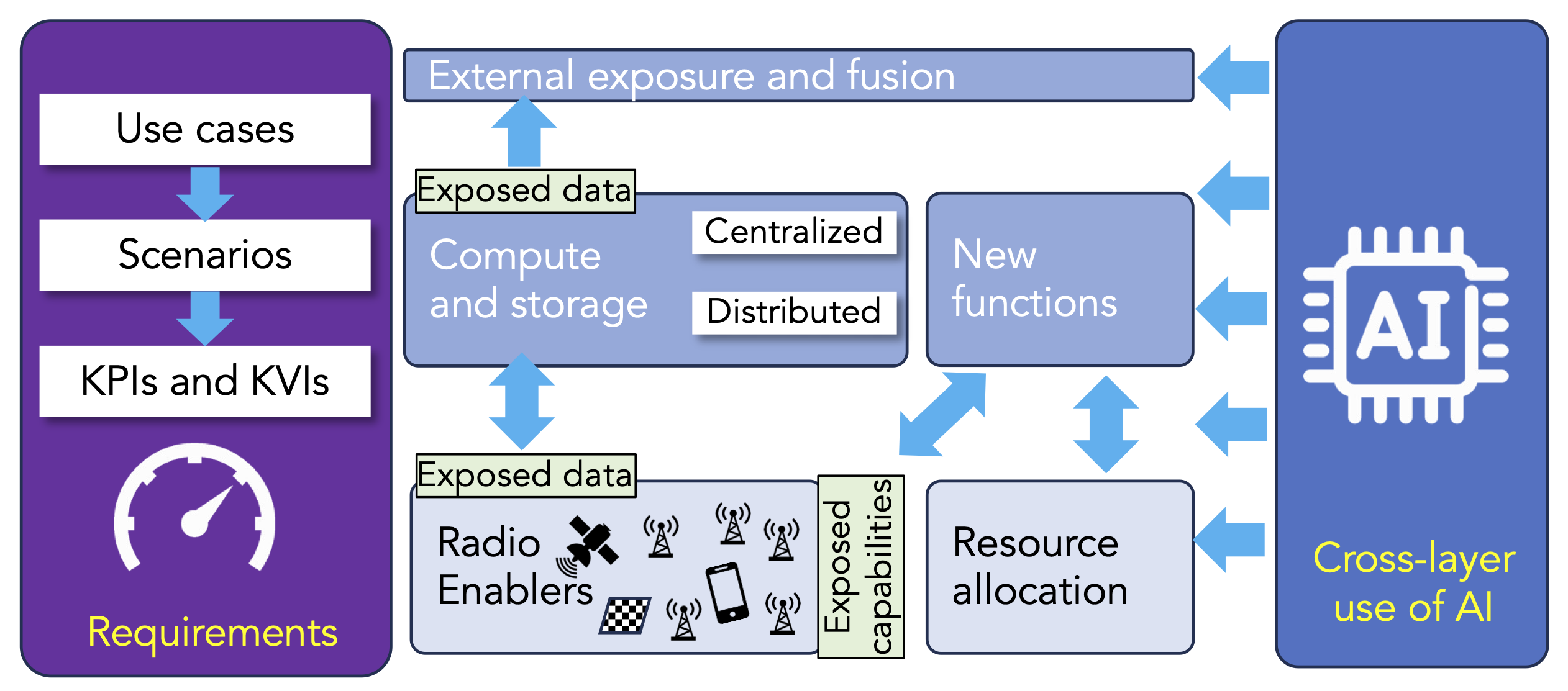}
    \caption{A cross-layer view of \ac{JCAS}, relating the radio enablers (NTN, RIS, D-MIMO) and resource allocation with compute and storage, new functions, and AI, to provide external exposure in support of the 6G radio scenarios.}
    \label{fig:crosslayerview}\vspace{-5mm}
\end{figure}
\subsubsection{Trustworthy JCAS}
 Due to its differentiation from the classical communication functionality and the sensitive nature of the produced information, this awareness of the physical environment introduced by \ac{JCAS} leads to new challenges and limitations. More specifically, the sensing functionality may produce information that may belong to the private domain of an entity. Also, in many use cases, it may trigger procedures that create changes in the actual physical environment. Thus,  the successful introduction of \ac{JCAS} heavily relies on the level of trustworthiness achieved by the new functionality. We propose to measure the level of trustworthiness by evaluating the: i) reliability, ii) security, and iii) ethical operation of the new functionality.
 \begin{itemize}
     \item \emph{Reliability:}  In a sensing procedure, this component of trustworthiness relates to how reliable are the obtained results in terms of accuracy for a given \ac{QoS}. In particular, it evaluates the proximity of a sensing measurement to the actual ground truth. Reliability is enhanced with the efficient allocation of the available time, frequency, and space resources along with the use of sophisticated signal processing algorithms for the suppression of degrading phenomena, such as clutter.
     \item \emph{Security:} This component of trustworthiness is connected with the robustness of a sensing procedure against malicious attacks that aim to degrade or alternate a sensing observation. The security of a sensing procedure relies on the successful identification and characterization of an ongoing attack combined with the use of countermeasure algorithms which minimize the effect of this ongoing attack. For example, the effective detection and localization of jamming nodes~\cite{9569361} can enable the use of spatial filtering methods along with cryptographic resource allocation protocols.
     \item \emph{Ethical operation:} As a sensing procedure may have the capacity to create information, that is connected with the private domain of an entity, it is important to utilize procedures that ensure the privacy of sensed entities. In addition, a producer of sensing data, such as the network, needs to ensure that the collected data are handled with care and protected against unauthorized use.
 \end{itemize}

\subsection{The Need for a Cross-Layer Perspective}
It becomes clear that \ac{JCAS} will have to rely on enablers from the radio side (to generate the necessary measurements) as well as from higher layers (to provide the functions and computational/storage resources). Such a cross-layer perspective is at the core of Hexa-X-II and is visualized in Fig.~\ref{fig:crosslayerview}. The right part of the figure shows transformative integration of \ac{AI},  
wherein AI orchestrates network-wide resource optimization for specific tasks rather than optimizing individual components. 
Thus, the integration of AI into JCAS naturally emerges from the inherent contradiction in optimal resource allocation between sensing and communications.
AI-powered JCAS holds the potential to obtain optimal task-specific trade-offs, and significantly enhance network sustainability and performance. 
In 3GPP Release 18 \cite{3gppai2023}, the integration of \ac{ML} and \ac{AI} was explored, where \ac{AI} replaced traditional model-based components, aiming to enhance the 5G \ac{NR} air interface. 6G proposes a holistic approach instead, enabling AI to further design and manage physical and medium access control resources \cite{hoydis2021toward}. AI is envisioned to seamlessly integrate physical and higher layer interactions, dynamically adapting networking resources based on radio side measurements.

\section{6G Radio for JCAS}

In this section, expanding Fig.~\ref{fig:crosslayerview},  we discuss \ac{JCAS} from the radio perspective, covering the physical deployments, resources, as well as enabling technologies. 
\begin{figure}
    \centering
    \includegraphics[width=0.9\linewidth]{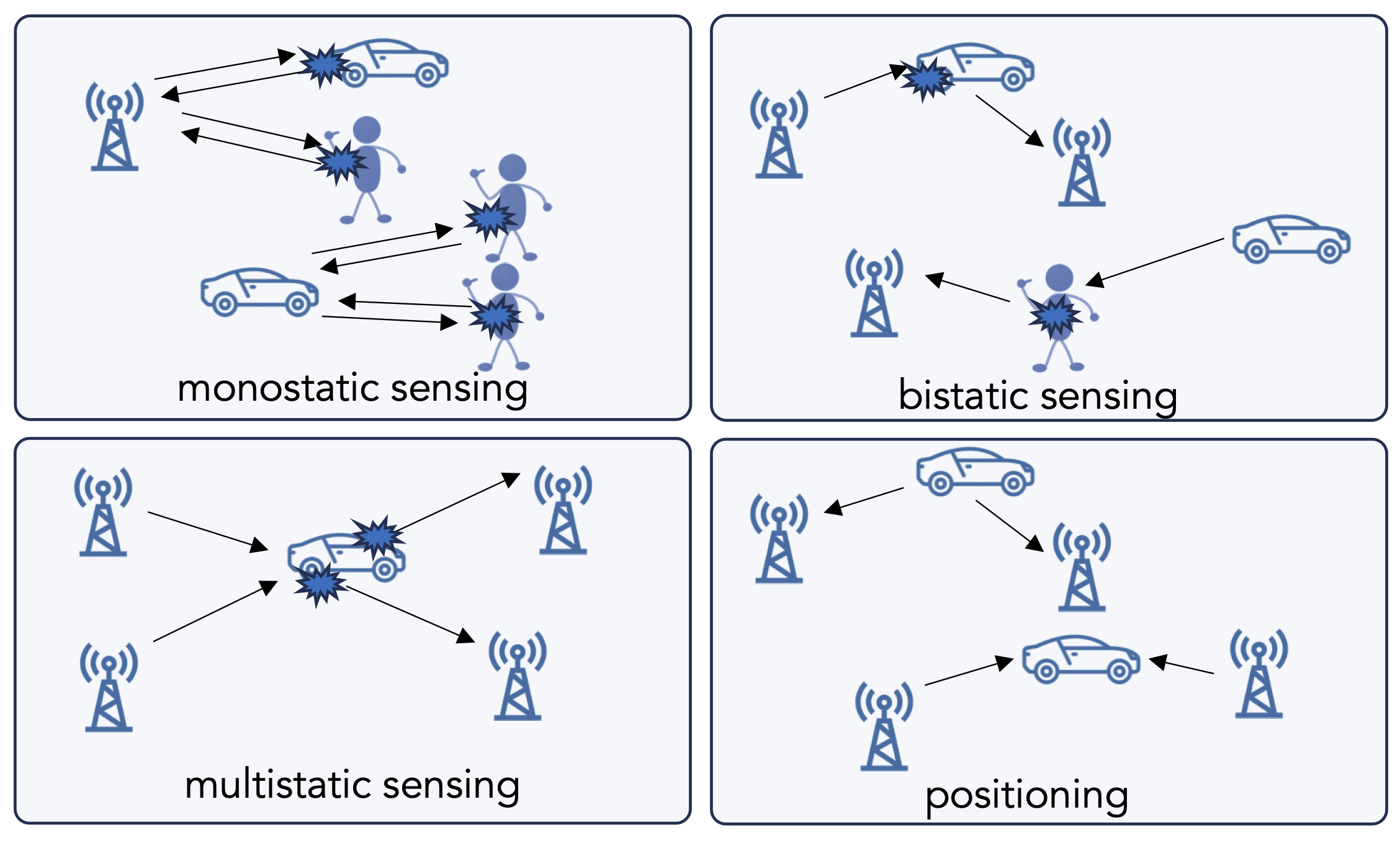}
    \caption{\ac{JCAS} deployments considered in Hexa-X-II. Each of the deployments can operate in uplink, downlink, and sidelink. Positioning can be seen as a service using bistatic sensing. }
    \label{fig:deployments}\vspace{-5mm}
\end{figure}
\subsection{Deployment Alternatives and Radio Resource Allocation}

Just as conventional positioning that can be formed in \ac{UL}, \ac{DL}, or even \ac{SL} transmission, the same is possible with sensing (see Fig.~\ref{fig:deployments}). 
Sensing includes monostatic sensing, where transmitter and receiver are co-located, sharing an oscillator, at the expense of requiring a full-duplex transceiver or dedicated sensing signals (e.g., pulse signals). Moreover, sensing between \acp{BS} is also foreseen in the form of multistatic sensing, where one or more \acp{BS} receive and process the signals from one or more transmitting \acp{BS}. Some of the sensing modes can even occur concurrently, such as monostatic and bistatic sensing, which provides new challenges for integration.

The concurrent operation of different sensing modes and communication over the same spectrum requires careful radio resource allocation.  Resource allocation is performed to avoid interference and to ensure high-quality communication and sensing functions, in support of the radio scenarios from Section \ref{sec:radioscenarios}. Resource allocation in the \ac{JCAS} context provides immense degrees of freedom since user data can be re-used for sensing at no cost in radio resources. Alternatively, dedicated pilots can be used to ensure a guaranteed sensing performance. 
Sensing can be performed sparsely in time, due to the slow mobility of objects (within the channel coherence time), and may rely on several \acp{BS} to provide a wide field of view.

\subsection{Enabling Technologies}
Within the vision for 6G to improve sustainability and inclusiveness, new enabling technologies will be employed to provide the necessary developments for JCAS. Some of these are \acp{NTN}, \acp{RIS}, \ac{D-MIMO}, Adaptive Waveform Design, Intelligent Signal Processing, etc.

\begin{itemize}
    \item \emph{NTNs}
integrate low Earth orbit (LEO) and geosynchronous Earth orbit (GEO) satellite constellations as well as high altitude platform stations (HAPS) to expand the reach of localization services to remote and rural areas \cite{5g_americas_5g_2020}. 
Localization can be based on NTN alone, or integrated with terrestrial networks (TNs), reducing the vertical dilution of precision (DoP). 
Although the integration of NTNs with emerging technologies like RIS is currently understudied, it is expected to enhance positioning accuracy considerably for single LEO satellite positioning \cite{wang_beamforming_2023}. 

\item \emph{\Acf{RIS}}
comprises passive meta-material elements that actively control the propagation environment, extending the reach of 6G TNs and NTNs deployed in dense/deep urban environments \cite{bjornson_reconfigurable_2022}. RIS enhances coverage and robustness by serving as supplementary anchors, offering \ac{LoS}  measurements for positioning  \cite{keykhosravi_leveraging_2023} and sensing  \cite{chepuri2023integrated}, with very high angular resolution.

\item \emph{D-MIMO:}
In contrast to \ac{NTN} and \ac{RIS}, which are mainly geared towards improving coverage with low cost, \ac{D-MIMO} is suited for scenarios with extreme sensing requirements. By extending the conventional resolution in range and angle domain, with an extremely large aperture from phase-coherent access points (APs), and high \ac{LoS} probability to a small number of APs, 
highly accurate localization and sensing becomes possible \cite{sakhnini2022near}. 

\end{itemize}

\section{6G compute and protocols for JCAS} \label{sec:higher-layer-view}
The evolution of the network towards \ac{JCAS}, and the corresponding radio enablers, will also require changes at higher layers as well (see again Fig.~\ref{fig:crosslayerview}). These changes relate to compute and storage, centralized and distributed processing, data exposure and fusion, as well as new functions,  and must be designed to match the radio enablers. 

\subsection{Compute and Storage}
Sensing and localization will likely generate large amounts of data with characteristics distinctly different from the existing user and control plane data. These expanded data volumes, or beyond-communication data, will have to be managed by the network or fused at various network locations, like access points, for efficient and coherent processing \cite{QCK22a}. Hence, appropriate design measures must be implemented to ensure realistic scaling that does not compromise either the delivery or the integrity of standard control plane data. These measures might encompass interfaces for transferring the sensing data to a new data plane and to external entities, as required. 
{The examination of the trade-off between various storage options is crucial for JCAS. \cite{ZDL23a} Data storage can be accomplished by various methods, including no-storage, in-memory storage, and external database storage. Concerning communication overhead, 
both no-storage and in-memory storage options are well-suited for applications that require quick access and processing of data. 
The utilization of external database storage can potentially result in increased communication overhead as a result of the need to retrieve and update stored data. Additionally, the selection of storage medium has a direct influence on the accessibility of historical information. Both no-storage and in-memory storage options are well-suited for real-time applications, as they allow for quick access and processing of data. On the other hand, external database storage provides the capability to analyze historical data, which is crucial for identifying long-term trends and making informed decisions.}

The ramifications of the data increase extend to the computational domain, thereby necessitating a reevaluation of the system’s architecture  \cite{de2019foundations}. 
Rather than merely managing data, we must also efficiently allocate the computational tasks related to the data \cite{zhu2023pushing,QCK22a}. As such, when a device or network node decides to offload a computation, it will have to discover and select the candidate compute nodes, capable of performing the requested computation while satisfying the associated KPIs. To efficiently perform the processing (compute) node selection, it is required to precisely define the parameters exchanged during the discovery and localization procedure, including processing (computing) capabilities of network and/or device nodes and requirements, such as latency and computational load.

\subsection{Centralized and Distributed Processing}

The architectural blueprint of beyond communication services, inclusive of centralized and distributed sensing, significantly influences compute and storage. Centralized configurations predominantly engage pre-configured nodes and procedures, thereby ensuring a relatively static and controlled environment. In contrast, distributed sensing introduces layers of complexity, enabling a dynamic environment where sensing data is both generated and utilized across diverse pre-processing stages, ranging from raw I/Q data to semantically enriched information \cite{CZC18a}. The convergence of communication, computing, and sensing will bring stringent prerequisites on latency, privacy/security, power consumption, and data accuracy \cite{TBB22a}. 
Therefore, it is necessary to introduce the corresponding novel architectural enablers,  including additions and/or modifications of network protocols and procedures. This imposes several challenges in connection establishment procedures that must be addressed, such as discovery, synchronization, and coordination of the multiple sensing and computing nodes.

\subsection{Exposure and Fusion}
The cross-layer nature of sensing functionality is evident: high-layer applications require geometrical contextual information derived from physical layer radio measurements, and on-demand sensing necessitates provisioning through specific high-layer network interfaces, adhering to defined resolution, reliability, and performance metrics.
The radio data at the physical layer (represented as raw I/Q samples or as detections with delay/angle/Doppler), undergoes refinement and compression before being relayed to the application \cite{TBB22a,ZDL23a}. 
Architectural considerations are key in deciding the location and timing of fusion, processing, and inference within the network architecture. To ensure proper scaling and manageable data volumes, it is crucial that data compression is done near the measuring node, especially if the node is a \ac{UE},  to avoid excessive raw I/Q data transmission \cite{QCZ20a,CZC18a}. Conversely, preserving the usefulness of sensing information across network layers is vital for optimal accuracy and resolution. Ideally, raw radio measurements from different locations would converge at a central unit for fusion and inference into geometric values. Nonetheless, practical scenarios, especially with massive sensing functionality usage, pose challenges in balancing sensing performance with network architectural capacity.

{The exposure of data is a critical component of the JCAS system, enabling network entities to have access to crucial sources of data. The significance of exposure grows as the volume of data generated and utilized for JCAS increases. Firstly, trust distinction is crucial when sharing data outside of the network domain (i.e., with third-party applications). The increasing interplay between networks and applications, along with the utilization of data from diverse sources, has the potential to impact the exposure framework. The implementation of methods for exposure controls is of utmost importance, as it aligns with the need to adhere to data requirements and accommodate diverse trust relationships. It is imperative to monitor the degree to which third-party applications can access these data. Secondly, the increase in data exposure through APIs has the potential to cause significant spikes in traffic, hence requiring the optimization of exposure performance. Meanwhile, while the goal is to make these data accessible to nodes wanting to utilize them, at the same time exposure of the collected and aggregated data needs a strong authentication and authorization mechanism to ensure that security and privacy will not be breached \cite{TBB22a}. Finally, it is imperative to consider the issue of latency in relation to data exposure \cite{STL23c}. The inclusion of data gathering, aggregation, cleaning, labeling, and other preparatory procedures may result in an extended duration between the initial data collection process and the point at which the data becomes usable and exposed.}

\begin{figure}
    \centering
   \includegraphics[width=1\columnwidth]{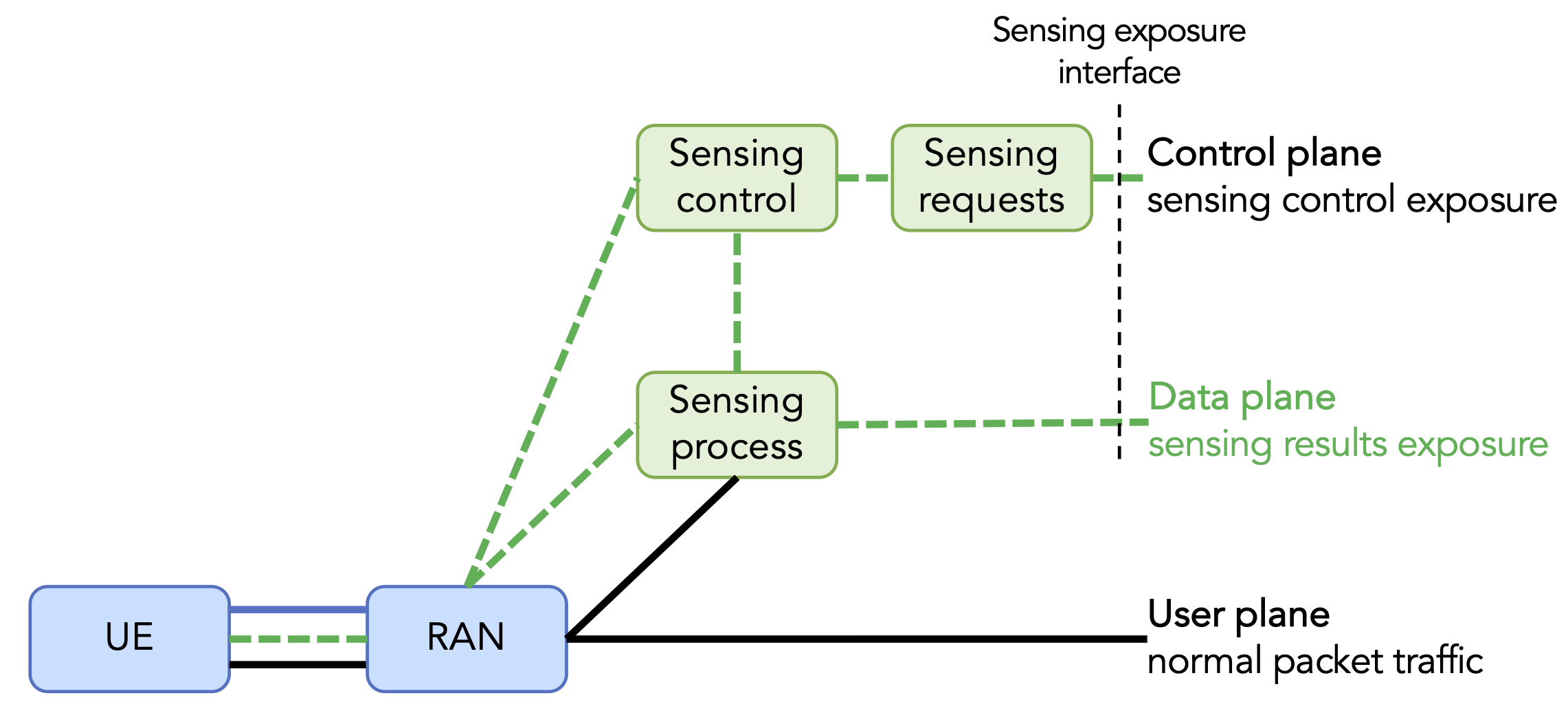}
    \caption{Logical nodes needed for sensing and how they connect to the architecture. The SeMF initiates the configuration of RAN for measurements on the control plane. Measurement data is forwarded from the RAN to the SPF over the data plane. Modified from~\cite{wanstedt2023}.} \vspace{-5mm}
    \label{fig:nodes}
\end{figure}
\subsection{New Network Functions}

The 6G inherent support for \ac{JCAS} necessitates certain network functions and architectural updates for the orchestration of the sensing procedures \cite{ZDL23a,STL23c}. Fig.~\ref{fig:nodes} outlines fundamental functionalities and presents the responsibilities of \acp{NF}, which are logical constructs that could be integrated into existing \acp{NF} or could be part of a dedicated Sensing Management Function (SeMF).
The sensing process begins with an application request (e.g., inquiring if an object exists within a specified area) received by the SeMF. The request includes an area, what type of reply that the application expects and, consequently, the complexity of the request can vary. Optionally, an Authorization Function may vet the request before it reaches SeMF. The SeMF then compares the requested area with the current cellular deployment and a 3D map of the surroundings to identify suitable nodes, like base stations and devices, needed to fulfill the request. Node selection depends on deployment and request's requirements; a simpler request may be handled with a monostatic setup, while complex ones may require a bi- or multi-static setup. Favorable scenarios exist if base stations have \ac{LoS} to the area; otherwise, \acp{UE} with \ac{LoS} might be activated. The processed request, with node configuration details, is used to initiate the configuration of the identified nodes needed for the measurement. The request and resulting configurations are forwarded over the control plane. 
Post-measurement outputs are sent to the SPF where the collected measurements are processed so as to derive the sensing result, according to the sensing request requirements. The measurements have characteristics similar to user plane data, however, there is no user to receive the data. Therefore, we propose a new \emph{data plane} for transporting raw I/Q data. Finally, The sensing result is transmitted to the requester over the user plane, releasing the sensing process.
Additionally, Fig.~\ref{fig:nodes} introduces a \emph{data plane} for transporting raw measurements, separate from the control and user planes, ensuring no user is connected to the raw measurement data.

\section{Conclusions}
The purpose of this paper was to emphasize the need for a cross-layer perspective to \ac{JCAS}, a topic addressed in the Hexa-X-II project. Starting from the use cases and radio scenarios, we first argued that new \acp{KPI} must be introduced to capture aspects related to sustainability and trustworthiness. We also provided an overview of the main enablers across the different layers. From this, it becomes clear that significant work remains to define appropriate interfaces and to support integration among the different layers, to realize \ac{JCAS} for 6G.

\balance 
\bibliographystyle{IEEEtran}
\bibliography{references.bib, ZoteroReferences}
\end{document}